\documentclass[preprint]{elsarticle}

\usepackage{hyperref,bm}
\usepackage{color}
\usepackage{ulem}

\usepackage{amsmath,amssymb}
\usepackage{booktabs}

\journal{Journal of \LaTeX\ Templates}

\begin{document}

\begin{frontmatter}

\title{
Stacking-dependent nonreciprocal transport in magnetic skyrmions
}

\author{Satoru Hayami}
\address{Graduate School of Science, Hokkaido University, Sapporo 060-0810, Japan}
\ead{hayami@phys.sci.hokudai.ac.jp}

\begin{abstract}
I theoretically propose an emergence of nonlinear nonreciprocal transport in centrosymmetric magnets with topological spin textures. 
By focusing on the stacking degree of freedom of the skyrmion crystals on a layered triangular lattice, I find that the ABC-stacking structure of the skyrmion crystals gives rise to out-of-plane nonreciprocal transport without the relativistic spin-orbit coupling. 
I show that such nonreciprocal transport is caused by an asymmetric band modulation due to the stacking structure, where the scalar chirality across the layers is a key ingredient. 
I also show an effective spin model to stabilize the ABC-stacking skyrmion structure. 
The present results indicate that the stacking structure of the skyrmion crystal becomes a source of further intriguing transport properties. 
\end{abstract}

\begin{keyword}
nonreciprocal transport, stacking degree of freedom, triangular lattice, magnetic skyrmion, spin-orbit coupling
\end{keyword}

\end{frontmatter}


\section{Introduction}

Magnetic metals have been extensively studied in both theory and experiment over the years, since they exhibit various fascinating physical phenomena that arise from an intimate interplay between the charge and spin degrees of freedom in electrons. 
Especially, magnetic phase transitions often lead to a drastic change in transport properties owing to the change in the electronic band structures and the lattice symmetry. 
One of the examples is colossal magnetoresistance, where there is an extreme change in longitudinal electrical conductivity under a ferromagnetic-to-paramagnetic phase transition~\cite{baibich1988giant,ramirez1997colossal,tokura1999colossal,Tokura200005,dagotto2001colossal}. 
Another example is the anomalous Hall effect caused by the spin Berry phase mechanism, where a transverse electric voltage occurs perpendicular to an input electric current and a magnetization without an external magnetic field~\cite{Loss_PhysRevB.45.13544, Ye_PhysRevLett.83.3737, Haldane_PhysRevLett.93.206602, Nagaosa_RevModPhys.82.1539, Xiao_RevModPhys.82.1959,Zhang_PhysRevB.101.024420}. 
The latter has been extensively investigated in ferromagnets~\cite{Karplus_PhysRev.95.1154,smit1958spontaneous,Maranzana_PhysRev.160.421,Berger_PhysRevB.2.4559,nozieres1973simple,Jungwirth_PhysRevLett.88.207208}, collinear antiferromagnets~\cite{Solovyev_PhysRevB.55.8060,Sivadas_PhysRevLett.117.267203,vsmejkal2020crystal}, noncollinear antiferromagnets~\cite{nakatsuji2015large,Suzuki_PhysRevB.95.094406}, and noncoplanar antiferromagnets~\cite{Ohgushi_PhysRevB.62.R6065,tatara2002chirality,Shindou_PhysRevLett.87.116801,Martin_PhysRevLett.101.156402}. 
Furthermore, spin-dependent conductive phenomena, such as the spin current and the spin Hall effect, in antiferromagnets have attracted much attention, as they can be induced even without the relativistic spin-orbit coupling and net magnetization~\cite{zhang2018spin,naka2019spin,hayami2019momentum}. 

In the present study, I investigate nonlinear nonreciprocal conductive phenomena under magnetic orderings in centrosymmetric lattice systems in the absence of spin-orbit coupling. 
Recent theoretical studies have revealed that spin structures breaking both spatial inversion and time-reversal symmetries can cause asymmetric electronic band modulations~\cite{Hayami_PhysRevB.101.220403, Hayami_PhysRevB.102.144441, Eto_PhysRevLett.129.017201}, and nonreciprocal transport~\cite{Cheon_PhysRevB.98.184405, Hayami_PhysRevB.106.014420,hayami2022nonreciprocal} without the spin-orbit coupling.  
I further aim at exploring such nonlinear transport phenomena by targeting a magnetic skyrmion crystal (SkX), which is characterized by a periodic array of swirling noncoplanar spin textures~\cite{Bogdanov89, Bogdanov94, rossler2006spontaneous, Muhlbauer_2009skyrmion, yu2010real, nagaosa2013topological, jani2021antiferromagnetic}. 
Although nonlinear transport properties of the SkX have been investigated in noncentrosymmetric magnets~\cite{Seki_PhysRevB.93.235131,giordano2016spin,kanazawa2017noncentrosymmetric,tokura2018nonreciprocal,yokouchi2018current, Hoshino_PhysRevB.97.024413, Santos_PhysRevB.102.104401,seki2020propagation, Kravchuk_PhysRevB.102.220408, Shen_PhysRevB.105.014422} where the Dzyaloshinskii-Moriya interaction plays an important role~\cite{dzyaloshinsky1958thermodynamic,moriya1960anisotropic}, those in centrosymmetric systems have been less studied especially for the direction perpendicular to the two-dimensional SkX plane. 
Since recent experimental studies have revealed the appearance of the SkXs in centrosymmetric hexagonal~\cite{kurumaji2019skyrmion} and tetragonal~\cite{khanh2020nanometric} lattice systems, it is desired to examine nonlinear transport properties in such centrosymmetric SkX-hosting magnets. 

I theoretically propose nonlinear nonreciprocal transport of the SkX without relying on spin-orbit coupling by focusing on its stacking degree of freedom. 
I find that a different stacking of the two-dimensional triangular SkX leads to qualitatively different nonlinear transport properties; the ABC-stacking structure of the SkXs causes the asymmetric electronic band structure and results in the Drude-type nonlinear nonreciprocal transport, while their A-stacking and AB-stacking structures do not show such properties. 
I reveal that the local scalar chirality across the layers plays an important role at the microscopic level.
I also show that interlayer antiferromagnetic exchange interaction is necessary to realize the ABC-stacking SkX by performing the simulated annealing for an effective spin model with the momentum-resolved interaction on a trilayer triangular lattice. 
The present results indicate that the stacking degree of freedom of the SkX is another important degree of freedom to control the nonreciprocal transport properties driven by the noncoplanar spin textures, which will be useful for other spin textures~\cite{Shen_PhysRevB.98.134448, Shen_PhysRevLett.124.037202,ye2022generation} and future spintronic applications based on skyrmions~\cite{zhang2020magnetic,xiao2020spin,shu2022realization}. 

The paper is organized as follows. 
First, I show that the stacking degree of freedom of the SkX is related to the phase degree of freedom among the constituent multiple-$Q$ spiral waves in Sec.~\ref{sec: Stacking of skyrmions}. 
Next, I discuss the electronic band structure and nonlinear nonreciprocal transport depending on the stacking structure of the SkX in Sec.~\ref{sec: Nonreciprocal transport}. 
I show that the ABC-stacking structure of the SkX becomes a source of nonreciprocal transport. 
Then, I present a minimal spin model to stabilize the ABC-stacking SkX by performing the simulated annealing in Sec.~\ref{sec: Simulated annealing}.  
Finally, a summary is presented in Sec.~\ref{sec: Summary}.
In Appendix~\ref{sec: One-dimensional system}, I discuss a minimal spin configuration to induce the asymmetric band modulation on a one-dimensional chain.

\section{Stacking of skyrmions}
\label{sec: Stacking of skyrmions}

\begin{figure}[htb!]
\begin{center}
\includegraphics[width=1.0 \hsize ]{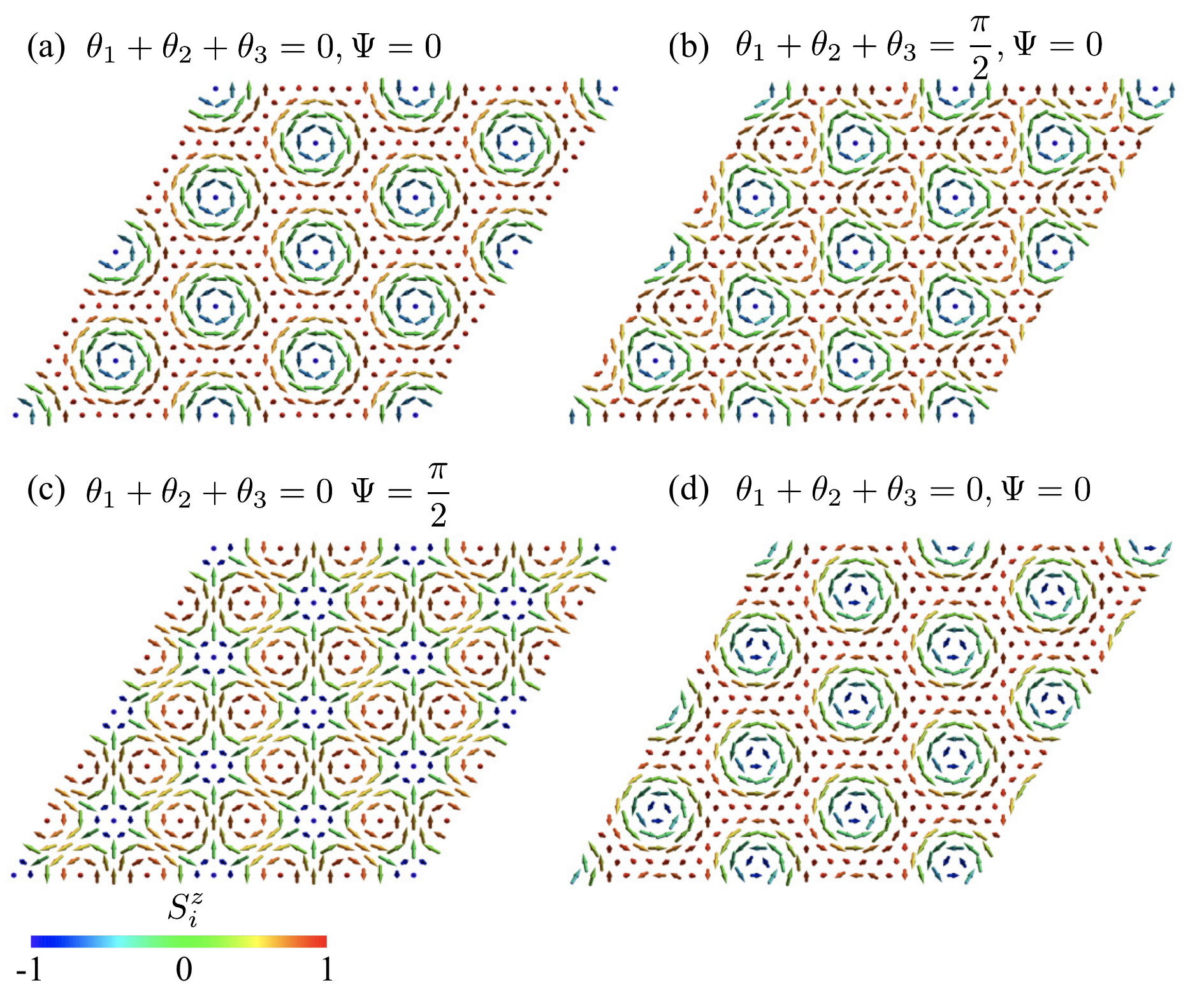} 
\caption{
\label{fig: ponti}
(a) Schematic spin configurations with different phases in Eq.~(\ref{eq: Si}): (a) $\theta_1=0$, $\theta_2=0$, $\theta_3=0$, and $\Psi=0$, (b) $\theta_1=\pi/2$, $\theta_2=0$, and $\theta_3=0$, and $\Psi=0$, (c) $\theta_1=0$, $\theta_2=0$, $\theta_3=0$, and $\Psi=\pi/2$, and (d) $\theta_1=\pi/3$, $\theta_2=-\pi/3$, $\theta_3=0$, and $\Psi=0$. 
The arrows show the spin moment and their color denotes the $z$-spin component; the red, blue, and green stand for the positive, negative, and zero components.
}
\end{center}
\end{figure}

\begin{figure}[htb!]
\begin{center}
\includegraphics[width=1.0 \hsize ]{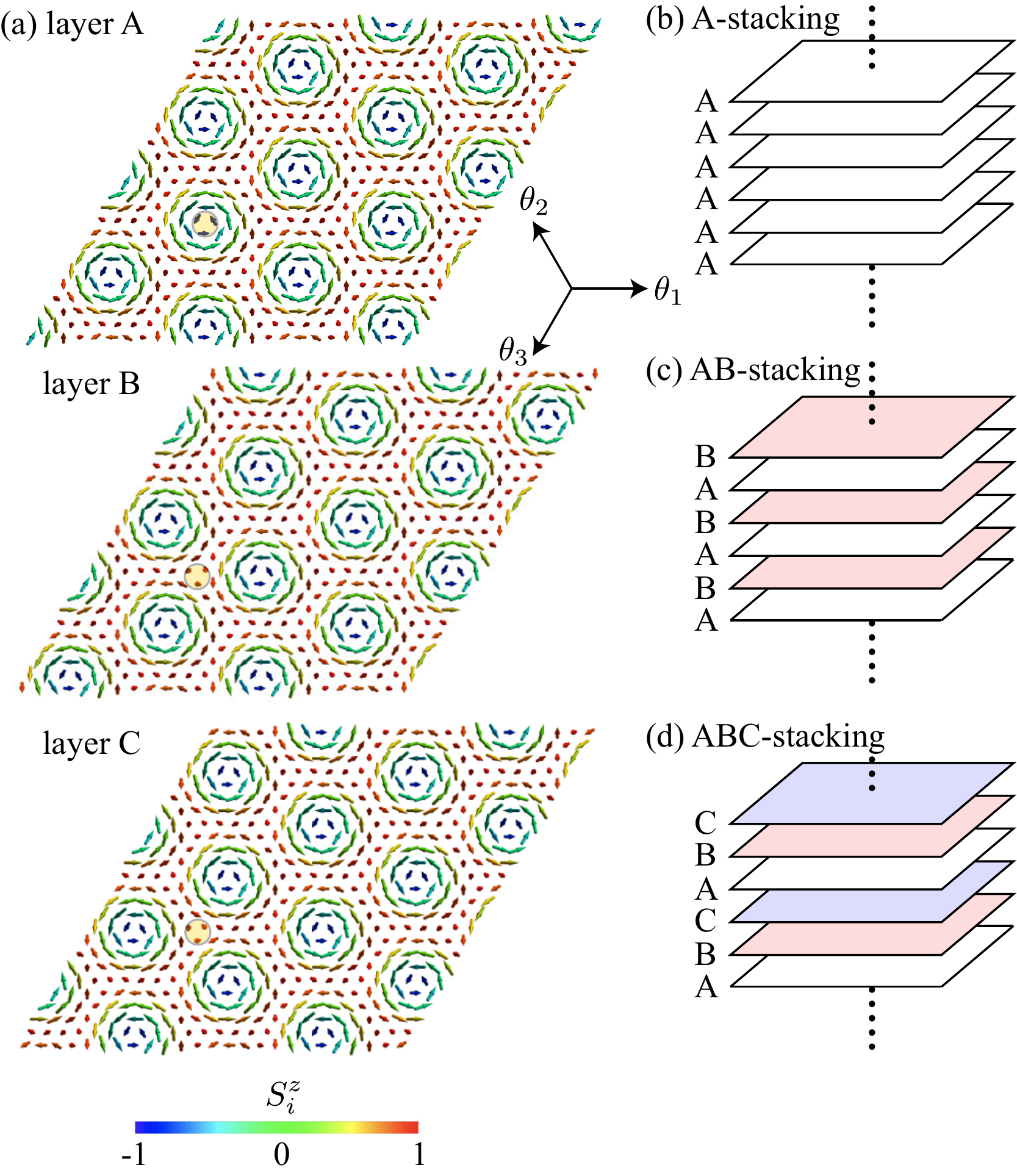} 
\caption{
\label{fig: setup}
(a) Three spin configurations for the skyrmion crystals (SkXs) with different phases in Eq.~(\ref{eq: Si}): (top panel) $\theta_1=\pi/3$, $\theta_2=-\pi/3$, and $\theta_3=0$, (middle panel) $\theta_1=-\pi$, $\theta_2=\pi/3$, and $\theta_3=2\pi/3$, and (bottom panel) $\theta_1=-\pi/3$, $\theta_2=-\pi$, $\theta_3=4\pi/3$. 
The arrows show the spin moment and their color denotes the $z$-spin component. 
The spin configuration in the top panel is the same as that in Fig.~\ref{fig: ponti}(d).
(b) Layered structures consisting of layers A--C in (a): (top panel) A-stacking structure, (middle panel) AB-stacking structure, and (bottom panel) ABC-stacking structure. 
}
\end{center}
\end{figure}

I first consider a spin configuration of the SkX on a two-dimensional triangular lattice with the lattice constant $a=1$; the lattice structure possesses the inversion center at the lattice points and the nearest-neighbor bond centers. 
The spin moment at site $i$, $\bm{S}_i = (S_i^x, S_i^y, S_i^z)$, is given by
\begin{eqnarray}
\label{eq: Si}
\bm{S}_i &=& \frac{\bm{m}_i}{|\bm{m}_i|}, \\
m^x_i &=& -\frac{\sqrt{3}}{2} \left[\sin(\mathcal{Q}_{2i} + \Psi)-\sin(\mathcal{Q}_{3i} + \Psi)\right], \\
m^y_i &=&\sin(\mathcal{Q}_{1i}+\Psi) -\frac{1}{2}\left[\sin(\mathcal{Q}_{2i} + \Psi)-\sin(\mathcal{Q}_{3i} + \Psi)\right], \\
m^z_i &=&\tilde{M}_z-a_z \left(\cos \mathcal{Q}_{1i} +\cos \mathcal{Q}_{2i} +\cos \mathcal{Q}_{3i} \right) , 
\end{eqnarray}
where $\mathcal{Q}_{\nu i}=\bm{Q}_\nu \cdot \bm{r}_i+\theta_\nu$ for $\nu=1$--3; $\bm{Q}_1=(\pi/3,0)$, $\bm{Q}_2=(-\pi/6,\sqrt{3}\pi/6)$, and $\bm{Q}_3=(-\pi/6,-\sqrt{3}\pi/6)$ represent the triple-$Q$ ordering vectors so as to satisfy $\bm{Q}_1+\bm{Q}_2+\bm{Q}_3=\bm{0}$, $\bm{r}_i$ is the position vector at site $i$, and $\theta_\nu$ and $\Psi$ are the phase degrees of freedom. 
The spin ansatz in Eq.~(\ref{eq: Si}) means that the spin configuration is characterized by a superposition of three proper-screw spiral waves along the $\bm{Q}_1$, $\bm{Q}_2$, and $\bm{Q}_3$ directions. 
Although I choose the specific vorticity ($+1$) and helicity ($\pi/2$) of the SkX~\cite{nagaosa2013topological}, it does not change the following results. 
The remaining $\tilde{M}_z$ and $a_z$ denote variational parameters depending on the model: The former stands for the uniform component of spins due to the superposition of $\bm{Q}_1+\bm{Q}_2+\bm{Q}_3=\bm{0}$ and the latter stands for the ellipticity of the spirals due to the inequivalence between the $xy$ and $z$ spin moments in the expression. 
I set $\tilde{M}_z=0.6$ and $a_z=0.7$. 
There are 48 spins in the magnetic unit cell, which is determined by the ordering vector $\bm{Q}_\nu$. 

The detailed spin textures of the SkX are affected by the internal phase degrees of freedom, $\theta_\nu$ and $\Psi$; $\theta_\nu$ represents the relative phase among the constituent waves along the $\bm{Q}_1$, $\bm{Q}_2$, and $\bm{Q}_3$ directions and $\Psi$ represents the relative phase between the $xy$ and $z$ component of spins. 
For example, $\sum_\nu \theta_\nu=0$ represents the SkX [Fig.~\ref{fig: ponti}(a)], while $\sum_\nu \theta_\nu=\pi/2$ represents the meron--antimeron crystal without a skyrmion number [Fig.~\ref{fig: ponti}(b)]~\cite{kurumaji2019skyrmion,hayami2021phase}; 
$\Psi=0$ ($\Psi=\pi/2$) describes the SkX with the skyrmion number of one (two)~\cite{yambe2021skyrmion}. 
The spin configuration of the SkX with the skyrmion number of two with $\sum_\nu \theta_\nu=0$ and $\Psi=\pi/2$ is shown in Fig.~\ref{fig: ponti}(c).
To investigate the nonlinear transport of the SkX with the skyrmion number of one, I take $\sum_\nu \theta_\nu=0$ and $\Psi=0$ in the following calculations. 
A different set of $\theta_\nu$ for fixed $\sum_\nu \theta_\nu=0$ describes the uniform translation of the SkX to the triangular lattice~\cite{Shimizu_PhysRevB.105.224405}; for $\theta_1=\theta_2=\theta_3=0$, the skyrmion core at $S_i^z=-1$ is located at the lattice site [Fig.~\ref{fig: ponti}(a)]. 
By taking $\theta_1=\pi/3$, $\theta_2=-\pi/3$, and $\theta_3=0$, the skyrmion crystal is translated so that the skyrmion core at $S_i^z=-1$ is located at the interstitial site, as shown in Fig.~\ref{fig: ponti}(d) and the top panel of Fig.~\ref{fig: setup}(a)~\cite{Hayami_PhysRevResearch.3.043158}. 
I also show the SkXs with different sets of $\theta_\nu$ as $(\theta_1,\theta_2,\theta_3)=(-\pi,\pi/3,2\pi/3)$ and $(-\pi/3,-\pi,4\pi/3)$ in the middle and bottom panels of Fig.~\ref{fig: setup}(a), respectively, where the skyrmion cores are located at different positions from those with $(\theta_1,\theta_2,\theta_3)=(\pi/3,-\pi/3,0)$. 
Hereafter, I denote the triangular-lattice layer with the SkX configurations with $(\theta_1,\theta_2,\theta_3)=(\pi/3,-\pi/3,0)$, $(-\pi,\pi/3,2\pi/3)$, and $(-\pi/3,-\pi,4\pi/3)$ as layer A, layer B, and layer C, respectively. 

I consider three types of the stacking structure of layers A--C on a centrosymmetric layered triangular lattice: A-stacking [Fig.~\ref{eq: Si}(b)], AB-stacking [Fig.~\ref{eq: Si}(c)], and ABC-stacking [Fig.~\ref{eq: Si}(d)] structures. 
Each layer is separated by $c=1$. 
There are $48$, $96$, and $144$ spins in the magnetic unit cell for the A-stacking, AB-stacking, and ABC-stacking structures, respectively.

\section{Nonreciprocal transport}
\label{sec: Nonreciprocal transport}

To examine the stacking-dependent nonlinear nonreciprocal transport under the SkX spin configuration in Eq.~(\ref{eq: Si}), I analyze one of the spin-charge coupled models, the Kondo lattice model, which consists of itinerant electrons and localized spins. 
The Kondo lattice Hamiltonian is given by 
\begin{equation}
\label{eq: Ham_KLM}
\mathcal{H}^{\rm KLM}= -\sum_{i, j,  \sigma} t_{ij} c^{\dagger}_{i\sigma}c_{j \sigma}
+J_{\rm K} \sum_{i} \bm{s}_i
\cdot \bm{S}_i, 
\end{equation}
where $c^{\dagger}_{i\sigma}$ ($c_{i \sigma}$) is a creation (annihilation) operator of an itinerant electron at site $i$ and spin $\sigma$, and $\bm{S}_i$ is a classical localized spin at site $i$ with $|\bm{S}_i|=1$. 
The first term in Eq.~(\ref{eq: Ham_KLM}) denotes the electron hoppings consisting of the nearest-neighbor intralayer hopping in the $xy$ plane $t_{ij}=t_{1}$ and the nearest-neighbor interlayer hopping along the $z$ direction $t_{ij}=t_{z}$; I fix $t_1=1$ as the energy unit in Secs.~\ref{sec: Electronic band structure} and \ref{sec: Drude-type nonlinear conductivity} and set $t_z=0.5$. 
The second term in Eq.~(\ref{eq: Ham_KLM}) represents the onsite exchange coupling between the itinerant electron spins $\bm{s}_i=(1/2)\sum_{\sigma,\sigma'}c^{\dagger}_{i\sigma} \bm{\sigma}_{\sigma \sigma'} c_{i \sigma'}$ and the localized spins $\bm{S}_i$ via the coupling constant $J_{\rm K}$; $\bm{\sigma}=(\sigma^x,\sigma^y,\sigma^z)$ is the vector of Pauli matrices. 
It is noted that the sign of $J_{\rm K}$ is irrelevant due to the classical nature of $\bm{S}_i$.

For the model in Eq.~(\ref{eq: Ham_KLM}), I discuss the electronic band structure in Sec.~\ref{sec: Electronic band structure} and the nonlinear nonreciprocal transport along the $z$ direction in Sec.~\ref{sec: Drude-type nonlinear conductivity} with an emphasis on the stacking structures. 
Although I suppose the SkX spin configurations in Eq.~(\ref{eq: Si}), I present a situation where such SkXs are stabilized in Sec.~\ref{sec: Simulated annealing}.

\subsection{Electronic band structure}
\label{sec: Electronic band structure}

\begin{figure}[t!]
\begin{center}
\includegraphics[width=1.0 \hsize ]{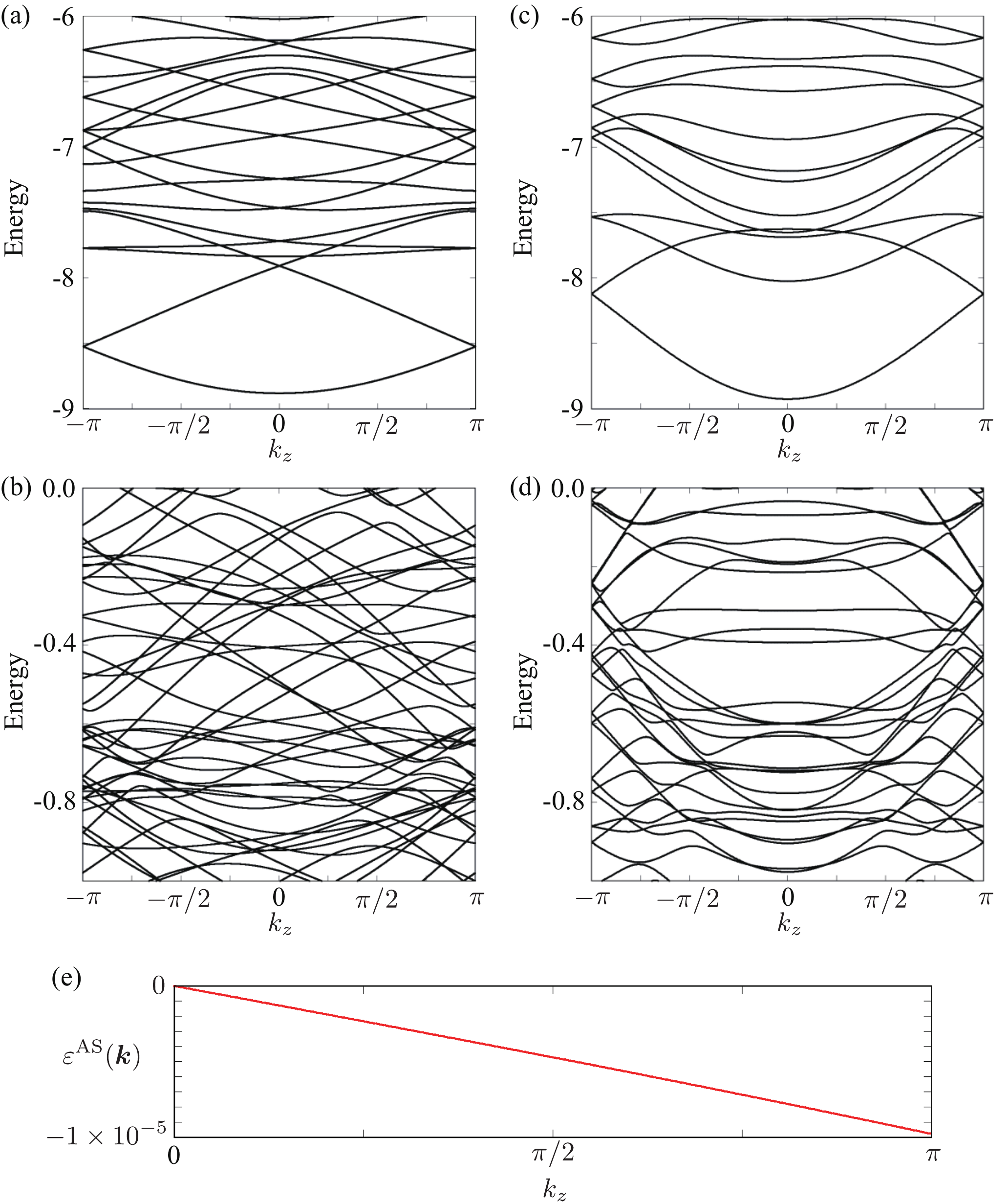} 
\caption{
\label{fig: band}
Electronic band dispersions for (a,b) the ABC-staking SkX structure and (c,d) the AB-stacking one at $J_{\rm K}=5$. 
The different energy ranges are plotted for (a) and (b) [(c) and (d)]. 
(e) Antisymmetric component of the energy dispersions for the lowest band $\varepsilon^{\rm AS}(\bm{k})$ corresponding to (a). 
}
\end{center}
\end{figure}

I first discuss the electronic band structure of the Kondo lattice model at $J_{\rm K}=5$ in Eq.~(\ref{eq: Ham_KLM}). 
Here and hereafter, I only consider the band dispersion along the $k_z$ direction, since I examine the nonlinear transport along the $z$ direction. 
As there are 288 bands in the magnetic unit cell in the case of the ABC-stacking structure, I present parts of the electronic band structures; the energy range from $-9$ to $-6$ in Fig.~\ref{fig: band}(a) and that from $-1$ to $0$ in Fig.~\ref{fig: band}(b). 
The energy of the overall band structure approximately ranges around from $-9$ to $6$. 
The bottom band in Fig.~\ref{fig: band}(a) corresponds to the lowest-energy band. 

By closely looking into the band structures in Figs.~\ref{fig: band}(a) and \ref{fig: band}(b), 
one finds that the band dispersions become asymmetric in terms of $k_z=0$, especially for the middle energy range in Fig.~\ref{fig: band}(b). 
Meanwhile, the band dispersion corresponding to the low-filling region in Fig.~\ref{fig: band}(a) seems to be symmetric. 
However, it is noted that there is a small antisymmetric component in the energy dispersions; I present $\varepsilon^{\rm AS}(\bm{k})=\varepsilon(-\bm{k})-\varepsilon(\bm{k})$ for the lowest-energy band in Fig.~\ref{fig: band}(e), where the antisymmetric modulation is small but nonzero. 
Thus, the ABC-stacking SkX exhibits the asymmetric electronic band structure along the $k_z$ direction. 

On the other hand, the AB-stacking and A-stacking SkXs do not show asymmetric band modulations; all the bands are symmetric in terms of $k_z$. 
As an example, I show the band structure in the AB-stacking SkX in Figs.~\ref{fig: band}(c) and \ref{fig: band}(d), which correspond to Figs.~\ref{fig: band}(a) and \ref{fig: band}(b), respectively.   
In contrast to the ABC-stacking structure, only the symmetric energy bands are obtained in the AB-stacking structure. 
This result indicates that the stacking structures of the SkX lead to qualitatively different electronic band structures, which result in different nonlinear nonreciprocal transport properties, as discussed in Sec.~\ref{sec: Drude-type nonlinear conductivity}. 

The appearance of the asymmetric band modulation indicates an active magnetic toroidal dipole moment under the ABC-stacking structure, as the symmetry of the $k_z$-linear band dispersion is the same as that of the magnetic toroidal dipole along the $z$ direction; both the spatial inversion and time-reversal symmetries are broken~\cite{Spaldin_0953-8984-20-43-434203, kopaev2009toroidal, Hayami_PhysRevB.98.165110}. 
More specifically, the ABC-stacking spin structure belongs to the same irreducible representation as the magnetic toroidal dipole along the $z$ direction; $A^-_{2u}$ under the hexagonal point group $D_{6{\rm h}}$~\cite{Hayami_PhysRevB.98.165110, Yatsushiro_PhysRevB.104.054412}.
Meanwhile, the origin of the asymmetric band deformation is different from the previous studies: The relativistic spin-orbit coupling is necessary to cause the asymmetric band deformation in the previous studies~\cite{Yatsushiro_PhysRevB.105.155157}, while the present mechanism based on noncoplanar spin textures in the SkXs does not rely on spin-orbit coupling.

Instead of relativistic spin-orbit coupling, the noncoplanar spin texture plays a role in inducing the asymmetric band modulation, since it gives rise to an effective magnetic field through the spin Berry phase mechanism, which results in similar physical phenomena in the presence of spin-orbit coupling~\cite{Loss_PhysRevB.45.13544, Ye_PhysRevLett.83.3737, Haldane_PhysRevLett.93.206602, Nagaosa_RevModPhys.82.1539, Xiao_RevModPhys.82.1959,Zhang_PhysRevB.101.024420}. 
Especially, I find that the local scalar chirality across three layers defined as 
\begin{equation}
\chi^{\rm layer}_i \equiv \bm{S}_i \cdot (\bm{S}_{i+\hat{z}}  \times \bm{S}_{i+2\hat{z}} ), 
\end{equation}
is the microscopic origin of the asymmetric band modulation; the subscript $i$ represents the site on layer A. 
Indeed, the system is not invariant for both spatial inversion and time-reversal symmetries for nonzero $\chi^{\rm layer}$. 
Moreover, this tendency is consistent with the result where there is no asymmetric band modulation in the AB-stacking and A-stacking SkXs, since $\chi^{\rm layer}=0$ owing to $\bm{S}_i = \bm{S}_{i+\hat{z}}$ or $\bm{S}_i = \bm{S}_{i+2\hat{z}}$ or $\bm{S}_{i+\hat{z}} = \bm{S}_{i+2\hat{z}}$. 
This result holds for other noncoplanar magnetic textures, such as a spiral state with a uniform magnetization, as discussed in the case of a one-dimensional chain in Appendix~\ref{sec: One-dimensional system}.

\subsection{Drude-type nonlinear conductivity}
\label{sec: Drude-type nonlinear conductivity}

\begin{figure}[t!]
\begin{center}
\includegraphics[width=0.5 \hsize ]{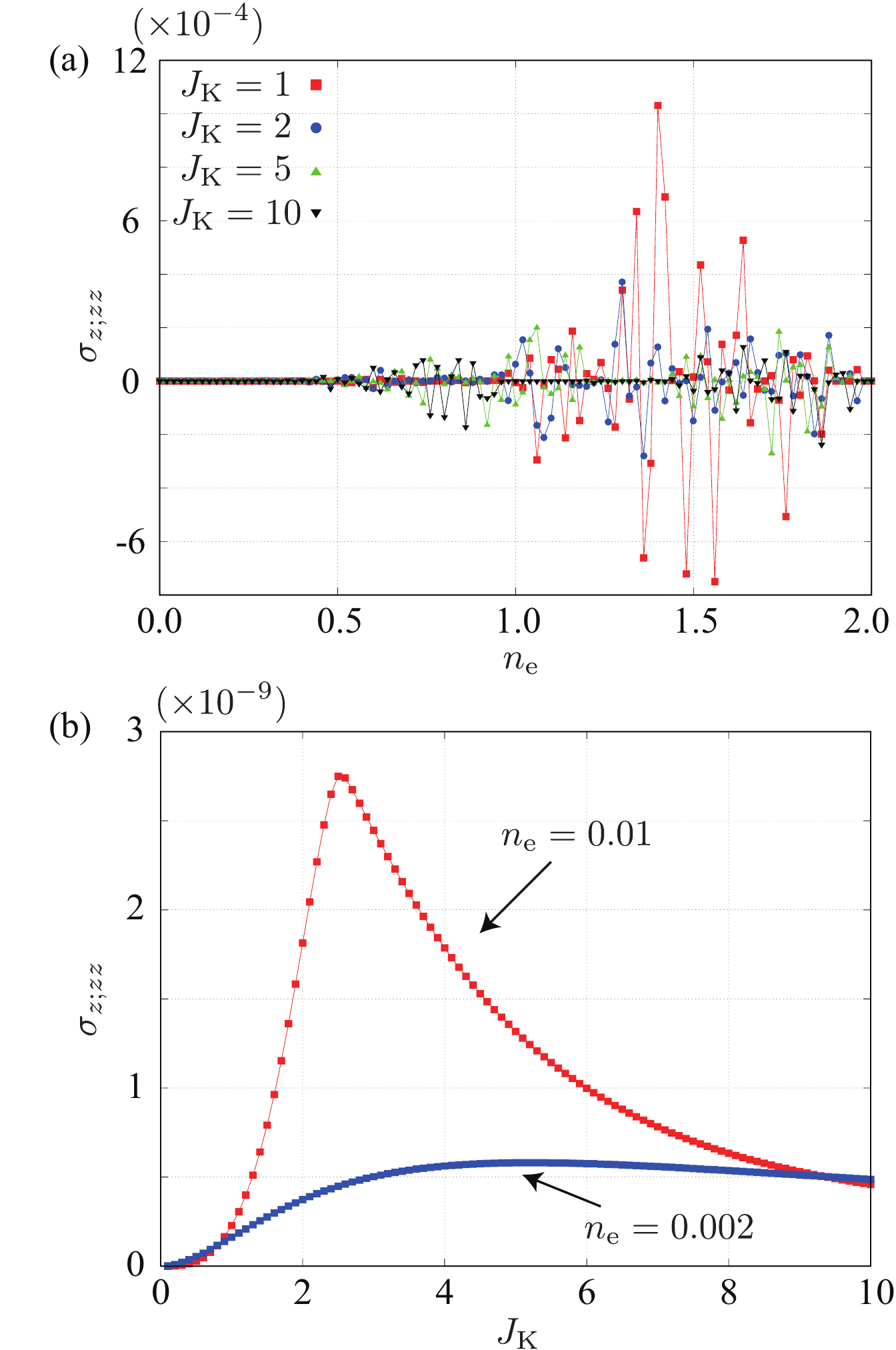} 
\caption{
\label{fig: conduc}
(a) $n_{\rm e}$ dependence of $\sigma_{z;zz}$ for $J_{\rm K}=1$, $2$, $5$, and $10$. 
(b) $J_{\rm K}$ dependence of $\sigma_{z;zz}$ for $n_{\rm e}=0.002$ and $0.01$. 
}
\end{center}
\end{figure}

Reflecting the asymmetric band modulation under the ABC-staking structure, the nonlinear nonreciprocal transport along the $z$ direction is expected. 
To demonstrate that, I calculate the Drude-type nonlinear conductivity $\sigma_{z;zz}$ in $J_{z}=\sigma_{z;zz} E_{z}E_{z}$, where $J_z$ and $E_z$ are the $z$ component of the electric current and electric field, respectively. 
The expression is derived from the second-order Kubo formula as 
\begin{equation}
\sigma_{z;zz}=-\frac{e^3 \tau^2}{2\hbar^3N_{\bm{k}}} \sum_{\bm{k},n}f_{n\bm{k}}\partial_{z}\partial_{z}\partial_{z}\varepsilon_{n\bm{k}}, 
\end{equation}
where $\varepsilon_{n\bm{k}}$ and $f_{n\bm{k}}$ are the eigenenergy and the Fermi distribution function with the band index $n$, respectively, which are obtained from the exact diagonalization under the ABC-stacking spin configuration. 
The coefficients $e$, $\tau$, $\hbar$, and $N_{\bm k}$ represent the electron charge, the relaxation time, the reduced Planck constant, and the number of supercells, respectively, where I set $e=\tau=\hbar=1$ and $N_{\bm{k}}=1\times 1 \times 12000$. 
It is noted that the $\tau$ dependence is different from that in the linear-order conductivity, the latter of which is proportional to $\tau$.
Hereafter, I fix the temperature $T=0.01$. 

Fig.~\ref{fig: conduc}(a) shows the behavior of $\sigma_{z;zz}$ against the filling $n_{\rm e}=(1/N)\sum_{i\sigma}c^{\dagger}_{i\sigma}c^{}_{i\sigma}$ with $N=144$ at $J_{\rm K}=1$, $2$, $5$, and $10$. 
As expected from the asymmetric band structure in Figs.~\ref{fig: band}(a) and \ref{fig: band}(b), $\sigma_{z;zz}$ becomes nonzero in the entire region of $n_{\rm e}$. 
There are two characteristic features in Fig.~\ref{fig: conduc}(a). 
One is the complicated behavior of $\sigma_{z;zz}$ including its sign change while changing $n_{\rm e}$, which might be attributed to the multi-band effect owing to a large number of sites in the magnetic unit cell. 
The other is small values of $\sigma_{z;zz}$ in the low-filling region for $n_{\rm e} \lesssim 0.5$ compared to that in the region for $n_{\rm e} \gtrsim 1$. 
It is noted that $\sigma_{z;zz}$ takes nonzero values in the low-filling region, as shown in the case of $n_{\rm e}=0.01$ and $0.002$ in Fig.~\ref{fig: conduc}(b). 
The reason why $\sigma_{z;zz}$ in the low-filling region is small is that the asymmetric band modulation in the low-filling region is smaller than that in the intermediate- and high-filling regions, as shown in Figs.~\ref{fig: band}(a) and \ref{fig: band}(b).

\section{Simulated annealing}
\label{sec: Simulated annealing}

\begin{figure}[htb!]
\begin{center}
\includegraphics[width=1.0 \hsize ]{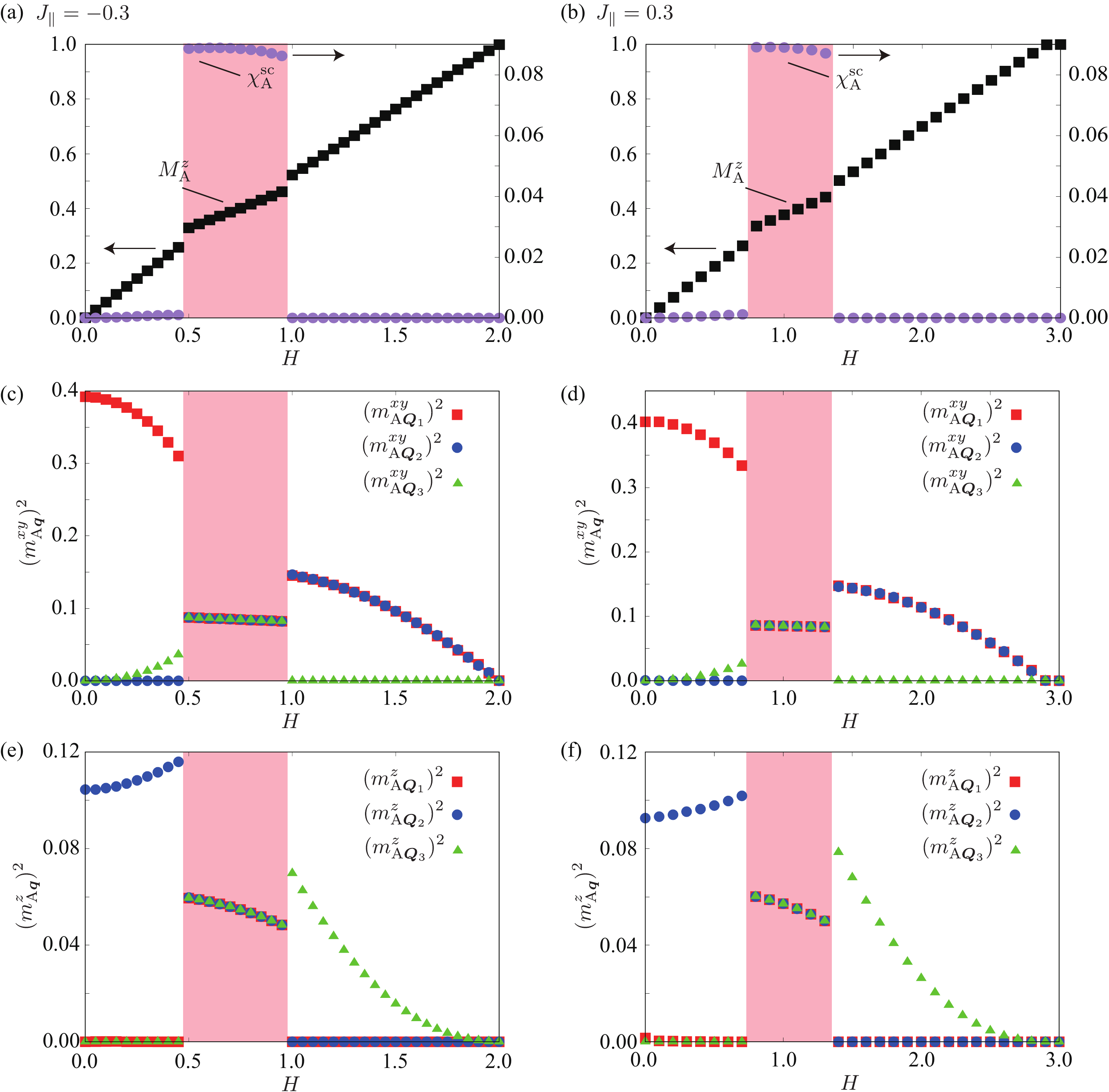} 
\caption{
\label{fig: mq}
$H$ dependence of (a,b) the uniform magnetization $M_{\rm A}^z$ and the scalar chirality $\chi_{\rm A}^{\rm sc}$, (c,d) the $xy$ component of magnetic moments $(m^{xy}_{{\rm A}\bm{q}})^2$, and (e,f) the $z$ component of magnetic moments $(m^{z}_{{\rm A}\bm{q}})^2$ for (a,c,e) $J_{\parallel}=-0.3$ and (b,d,f) $J_{\parallel}=0.3$. 
It is noted that the data for layers B and C are the same as those for layer A. 
The colored region corresponds to the SkX phase. 
}
\end{center}
\end{figure}

\begin{figure*}[htb!]
\begin{center}
\includegraphics[width=1.0 \hsize ]{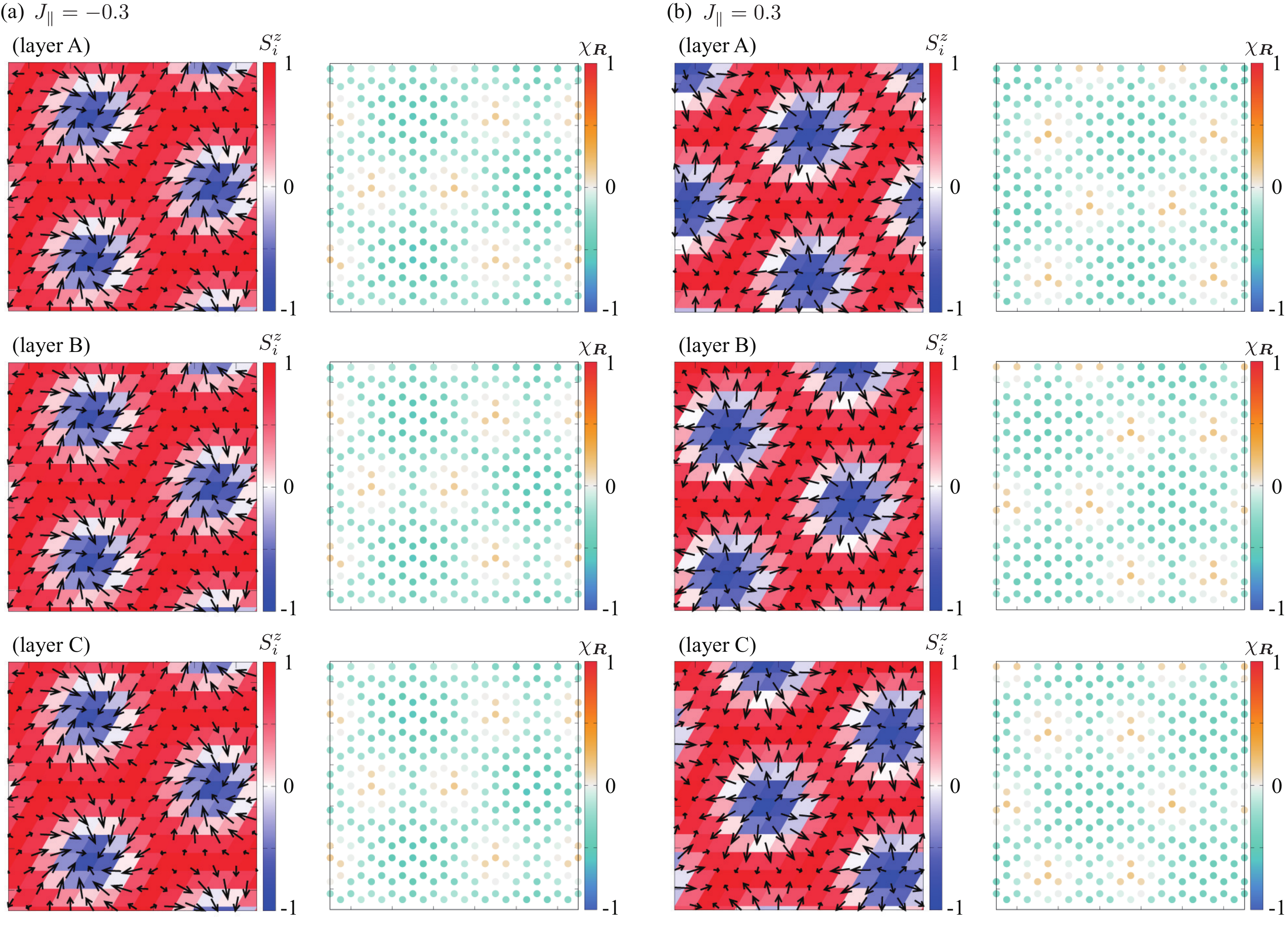} 
\caption{
\label{fig: spin}
Real-space spin (left panel) and scalar chirality (right panel) configurations of (a) the A-stacking SkX at $J_{\parallel}=-0.3$ and $H=0.8$ and (b) the ABC-stacking SkX at $J_{\parallel}=0.3$ and $H=0.8$ on layer A (top panel), layer B (middle panel), and layer C (bottom panel), which are obtained by the simulated annealing.  
In the left panels, the arrows represent the $xy$ spin and the color shows the $z$ spin. 
}
\end{center}
\end{figure*}

Finally, I show the important essence to stabilize the ABC-stacking SkX from the microscopic viewpoint. 
To this end, I consider the following effective spin model on a trilayer triangular lattice~\cite{Hayami_PhysRevB.105.184426}, whose Hamiltonian is given by 
\begin{eqnarray}
\label{eq: Ham}
\mathcal{H}&=&\sum_{\eta}\mathcal{H}^{\perp}_{\eta}+\mathcal{H}^{\parallel}+\mathcal{H}^{{\rm Z}}, \\
\label{eq: Ham_perp}
\tilde{\mathcal{H}}^{\perp}_\eta&=&  2\sum_{\nu}  \Big[- J \bm{S}^{(\eta)}_{\bm{Q}_{\nu}} \cdot \bm{S}^{(\eta)}_{-\bm{Q}_{\nu}}+ \frac{K}{N} ( \bm{S}^{(\eta)}_{\bm{Q}_{\nu}} \cdot \bm{S}^{(\eta)}_{-\bm{Q}_{\nu}})^2 \Big],  \\
\label{eq: Ham_parallel}
\mathcal{H}^{\parallel}&=& J_{\parallel} \sum_{i, \delta=\pm 1} \bm{S}_i \cdot \bm{S}_{i+\delta\hat{z}},\\
\label{eq: Ham_Zeeman}
\mathcal{H}^{{\rm Z}}&=&-H \sum_i S_i^z, 
\end{eqnarray}
where $\bm{S}^{(\eta)}_{\bm{Q}_{\nu}}$ is the Fourier transform of $\bm{S}_i$ with wave vector $\bm{Q}_\nu$ ($\nu$ is the index of the wave vectors) for layer $\eta=$ A, B, and C. 
The first term in Eq.~(\ref{eq: Ham}) represents the intralayer Hamiltonian $\tilde{\mathcal{H}}^{\perp}_\eta$ consisting of the bilinear exchange interaction with $J$ and the biquadratic interaction with $K$; both interactions are layer independent. 
The origin of $J$ and $K$ is the Fermi surface instability at $\bm{Q}_\nu$ in the weak coupling regime of the Kondo lattice model in Eq.~(\ref{eq: Ham_KLM}); $J$ and $K$ correspond to the lowest and second-lowest order of the expansion in terms of $J_{\rm K} $, respectively~\cite{Akagi_PhysRevLett.108.096401, Hayami_PhysRevB.90.060402, Hayami_PhysRevB.95.224424}.
I suppose that the nesting vectors are described by $\bm{Q}_1$, $\bm{Q}_2$, and $\bm{Q}_3$ in Sec.~\ref{sec: Stacking of skyrmions} for simplicity. 
It is noted that qualitatively similar results are obtained even when taking into account the interactions at other wave vectors unless the interactions at $\bm{Q}_\nu$ become smaller than those at other wave vectors. 
The second term in Eq.~(\ref{eq: Ham}) represents the interlayer Hamiltonian $\mathcal{H}^{\parallel}$ with the nearest-neighbor exchange interaction $J_{\parallel}$. 
The last term in Eq.~(\ref{eq: Ham}) represents the Zeeman Hamiltonian $\mathcal{H}^{{\rm Z}}$ under an external magnetic field with the magnitude of $H$. 

To examine the stacking tendency of the SkXs, I set the model parameters $J=1$ and $K=0.1$ so as to stabilize the SkX for $J_{\parallel}=0$, where $J$ is the energy unit of the model in Eq.~(\ref{eq: Ham}); the SkX is realized in the intermediate field~\cite{Hayami_PhysRevB.95.224424}. 
Then, I investigate the effect of $J_{\parallel}$ on the layered SkX structure, which lifts the energetic degeneracy among the A-, AB, and ABC-stacking SkXs. 
As I consider the trilayer system, I discuss the competition between the A-stacking and ABC-stacking SkXs for simplicity.

Fig.~\ref{fig: mq} shows the $H$ dependence of physical quantities in the case of the ferromagnetic $J_{\parallel}=-0.3$ in Figs.~\ref{fig: mq}(a), \ref{fig: mq}(c), and \ref{fig: mq}(e) and the antiferromagnetic $J_{\parallel}=0.3$ in Figs.~\ref{fig: mq}(b), \ref{fig: mq}(d), and \ref{fig: mq}(f), which are obtained by performing the simulated annealing following the manner in Ref.~\cite{Hayami_PhysRevB.95.224424}.  
I set the final temperature as $T=0.001$.
The data in Figs.~\ref{fig: mq}(a) and \ref{fig: mq}(b) represent the uniform magnetization per layer $M^z_\eta=(3/N)\sum_{i\in \eta}S^z_i$ and the scalar chirality per layer $\chi_\eta^{\rm sc}=(3/N)\sum_{\bm{R}\in \eta}\chi_{\bm{R}}$ with $\chi_{\bm{R}}=\bm{S}_i \cdot (\bm{S}_j \times \bm{S}_k)$, where $\bm{R}$ represents the center of the triangle plaquette with the vertices $i$, $j$, and $k$ in the counterclockwise order. 
The data in Figs.~\ref{fig: mq}(c) and \ref{fig: mq}(d) [Figs.~\ref{fig: mq}(e) and \ref{fig: mq}(f)] stand for the $xy$ ($z$) component of the magnetic moments at $\bm{Q}_\nu$ in each layer. 
Although I only show the data for layer A, the data for layer B and layer C are the same as those for layer A. 

The SkX phase with nonzero $\chi^{\rm sc}_{\rm A}$ and the triple-$Q$ magnetic moments appears in the intermediate-field region for both $J_{\parallel}=-0.3$ and $J_{\parallel}=0.3$, as shown in Fig.~\ref{fig: mq}. 
Meanwhile, the stacking structure is different for the ferromagnetic case and the antiferromagnetic case. 
I show the real-space spin and scalar chirality configurations of the SkX for $J_{\parallel}=-0.3$ and $H=0.8$ in Fig.~\ref{fig: spin}(a) and $J_{\parallel}=0.3$ and $H=0.8$ in Fig.~\ref{fig: spin}(b). 
As clearly shown in the figure, the A-stacking structure is stabilized for the ferromagnetic $J_{\parallel}$, while the ABC-stacking structure is realized for the antiferromagnetic $J_{\parallel}$. 
This is naturally understood from the energy gain by the interlayer exchange interaction for the A-stacking and ABC-stacking spin structures. 
When the interlayer exchange interaction is ferromagnetic ($J_{\parallel}<0$), the spins on the same $xy$ but different $z$ positions tend to align parallel. 
In other words, the A-stacking spin configuration is favored for $J_{\parallel}<0$. 
Similarly, when the interlayer exchange interaction is antiferromagnetic ($J_{\parallel}>0$), they tend to align antiparallel to gain the interlayer exchange interaction, which leads to the ABC-stacking SkX. 
Thus, one finds that the antiferromagnetic interlayer interaction assists the stabilization of the ABC-stacking SkX to exhibit nonlinear nonreciprocal transport.

\section{Summary}
\label{sec: Summary}

To summarize, I have investigated the nonlinear nonreciprocal transport of the SkX in centrosymmetric lattice systems without relativistic spin-orbit coupling by focusing on the stacking degree of freedom. 
I showed that the stacking degree of freedom in the SkX, which is closely related to the internal phase degree of freedom among the constituent waves, is another important degree of freedom to induce nonreciprocal transport along the layered direction. 
Indeed, I found that the ABC-stacking structure of the SkXs leads to asymmetric band modulation and nonlinear nonreciprocal transport, while the A-stacking and AB-stacking structures do not.  
Moreover, I presented a minimal model to realize the ABC-stacking SkX, where the antiferromagnetic interlayer interaction plays an important role.

As the nonlinear transport property along the out-of-plane direction is sensitive to the skyrmion alignment, it provides useful information to identify the real-space SkX stacking in the skyrmion-hosting materials under the centrosymmetric lattice structures. 
In addition, the local scalar chirality across the layers $\chi^{\rm layer}$ can be used as the indicator for arbitrary stacking patterns even with different spin configurations including antiferromagnetic SkX and other topological spin crystals. 
In this way, the present results will shed light on the importance of the stacking degree of freedom in centrosymmetric magnets with noncoplanar spin textures.

\appendix
\section{One-dimensional system}
\label{sec: One-dimensional system}

\begin{figure}[htb!]
\begin{center}
\includegraphics[width=0.6 \hsize ]{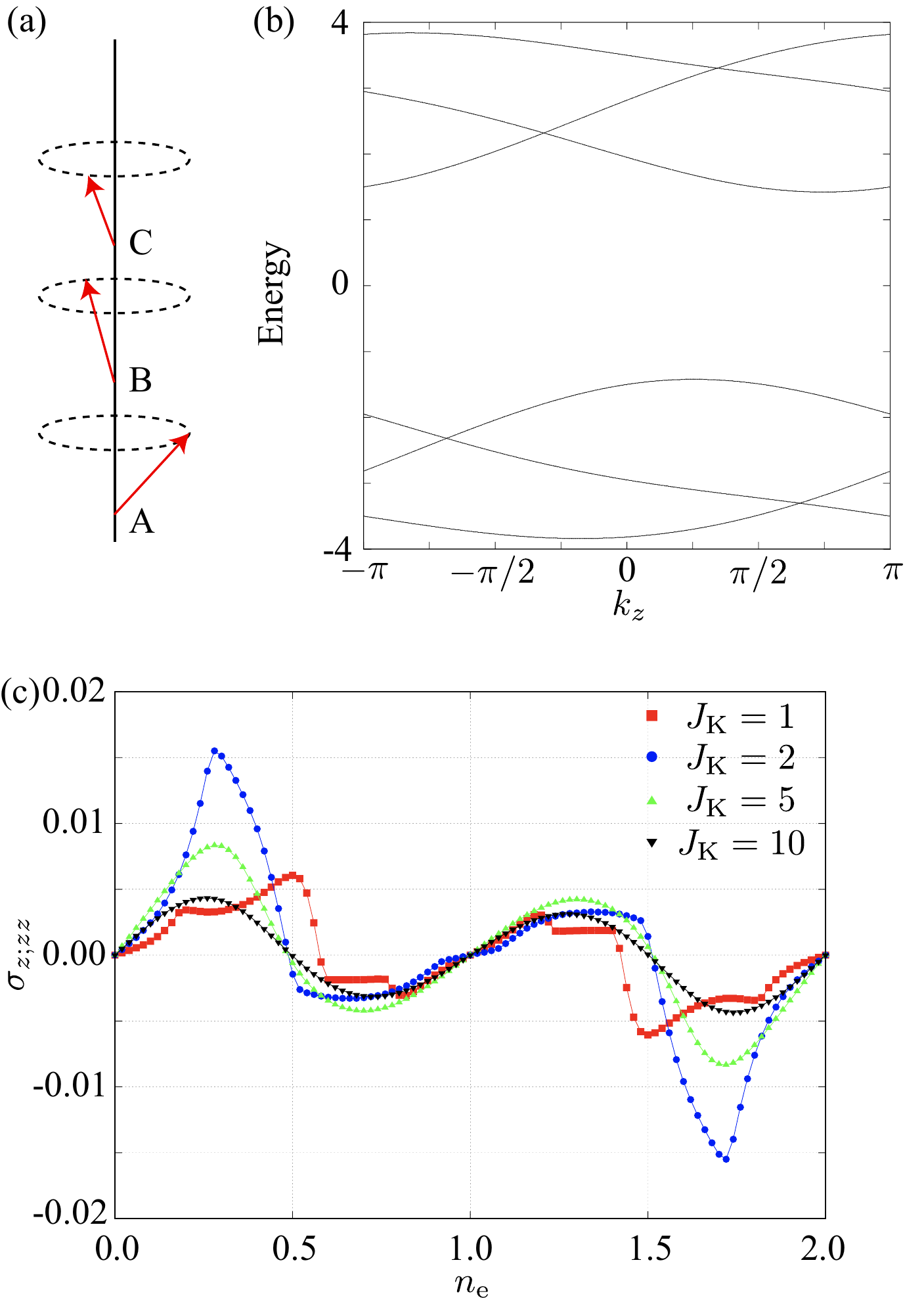} 
\caption{
\label{fig: app}
(a) Three-site umbrella spin configuration. 
(b) Electronic band dispersions for (a) at $J_{\rm K}=5$. 
(c) $n_{\rm e}$ dependence of $\sigma_{z;zz}$ for (a) at $J_{\rm K}=1$, $2$, $5$, and $10$. 
}
\end{center}
\end{figure}

In this Appendix, I present a minimal spin configuration to induce the asymmetric band modulation by considering a three-sublattice one-dimensional chain in Fig.~\ref{fig: app}(a).  
The three sublattices are denoted as A, B, and C. 
To consider the spin configuration with nonzero $\chi^{\rm layer}$, I consider an umbrella spin configuration in Fig.~\ref{fig: app}(a); $\bm{S}_{\rm A}=(\sin \theta, 0, \cos \theta)$, $\bm{S}_{\rm B}=(-\sin \theta/2, \sqrt{3}\sin \theta/2, \cos \theta)$, and $\bm{S}_{\rm C}=(-\sin \theta/2, -\sqrt{3}\sin \theta/2, \cos \theta)$ with $\theta=\cos^{-1}(1/3)$.

Figure~\ref{fig: app}(b) shows the electronic band structure under the umbrella spin configuration at $t_z=1$ and $J_{\rm K}=5$. 
The result clearly indicates that the asymmetric band modulation appears for $\chi^{\rm layer} \neq 0$, which is similar to the result in Fig.~\ref{fig: band}(a). 
Furthermore, no asymmetric band modulation occurs in the coplanar spin texture so that the three-sublattice spins lie on a plane or $\bm{S}_{\rm A} = \bm{S}_{\rm B}$ or $\bm{S}_{\rm B} = \bm{S}_{\rm C}$ or $\bm{S}_{\rm C} = \bm{S}_{\rm A}$. 
Reflecting the asymmetric band structure, the Drude-type nonlinear conductivity $\sigma_{z;zz}$ becomes nonzero irrespective of $J_{\rm K}$, as shown in Fig.~\ref{fig: app}(c). 
Thus, the umbrella-type noncoplanar spin configuration is also a candidate to exhibit the nonlinear nonreciprocal transport even without relativistic spin-orbit coupling.

This research was supported by JSPS KAKENHI Grants Numbers JP21H01037, JP22H04468, JP22H00101, JP22H01183, and by JST PRESTO (JPMJPR20L8). 
Parts of the numerical calculations were performed in the supercomputing systems in ISSP, the University of Tokyo.

\bibliography{ref}

\end{document}